\documentclass[conference]{IEEEtran}
\IEEEoverridecommandlockouts
\usepackage{cite}
\usepackage{amsmath,amssymb,amsfonts}
\usepackage{algorithmicx}
\usepackage{graphicx}
\usepackage{textcomp}
\usepackage{xcolor}
\usepackage{algorithm}
\usepackage{algpseudocode}

\DeclareMathOperator*{\blkdiag}{blkdiag}

\def\BibTeX{{\rm B\kern-.05em{\sc i\kern-.025em b}\kern-.08em
    T\kern-.1667em\lower.7ex\hbox{E}\kern-.125emX}}
\begin{document}

\title{Target Localization and Self-Calibration in a Multistatic Radar System
\thanks{The authors are with the School of Aeronautics and Astronautics, Purdue University, West Lafayette, IN, USA (e-mail: amusalla@purdue.edu; husheng@purdue.edu) \\This work was supported by the National Science Foundation under grant 2418106.}
}

\author{\IEEEauthorblockN{Ahmad Musallam, Husheng Li, Senior Member, IEEE}
}

\maketitle

\begin{abstract}
Target localization in a multistatic radar system, where multiple receivers cooperate to improve target positioning accuracy, has many applications, including cooperative simultaneous localization and mapping (SLAM) and autonomous robot networks. A key challenge in these applications is the uncertainty in the position and orientation (pose) of the radar receivers due to platform mobility. This work investigates the achievable improvements in both target localization and receiver pose estimation by deriving the Cramer-Rao lower bound (CRLB) for a multistatic radar system performing bistatic range and bearing measurements. We propose an alternating weighted least-squares algorithm that jointly optimizes target and receiver parameters. Monte Carlo simulations demonstrate that the algorithm performance approaches the CRLB for low to moderate noise levels.
\end{abstract}


\section{Introduction}
Multistatic radar systems have recently attracted attention \cite{10897834} because of their distributed and cooperative nature, which can provide spatial diversity and allow for better target localization. In many practical deployments of such systems, the radar receivers themselves are mobile platforms (e.g., vehicles, drones, or shipboard units) whose exact positions and orientations (poses) are not known. Cooperation in such systems is difficult because the uncertainty in the radar receiver poses leads to additional errors when trying to communicate the observed target position in a global reference frame with other receivers in order to perform data fusion. This is particularly an issue in multistatic radar systems, where it is desirable to change the receiver poses to illuminate different areas and improve the radar sensing geometry \cite{7946261}.

Joint target localization and receiver self-calibration represents a critical yet challenging task in multistatic radar systems. The difficulty stems from the fact that bistatic range and bearing measurements are highly nonlinear and very sensitive to errors in the pose of the receiver, which creates a highly nonconvex estimation problem. In addition, the measurement accuracy in multistatic radar systems is highly geometry-dependent, making it difficult to create a robust algorithm that efficiently solves the joint problem while dealing with noise from both measurements and the receiver position. 
\begin{figure}[ht]
    \centering
    \includegraphics[width=0.9\linewidth]{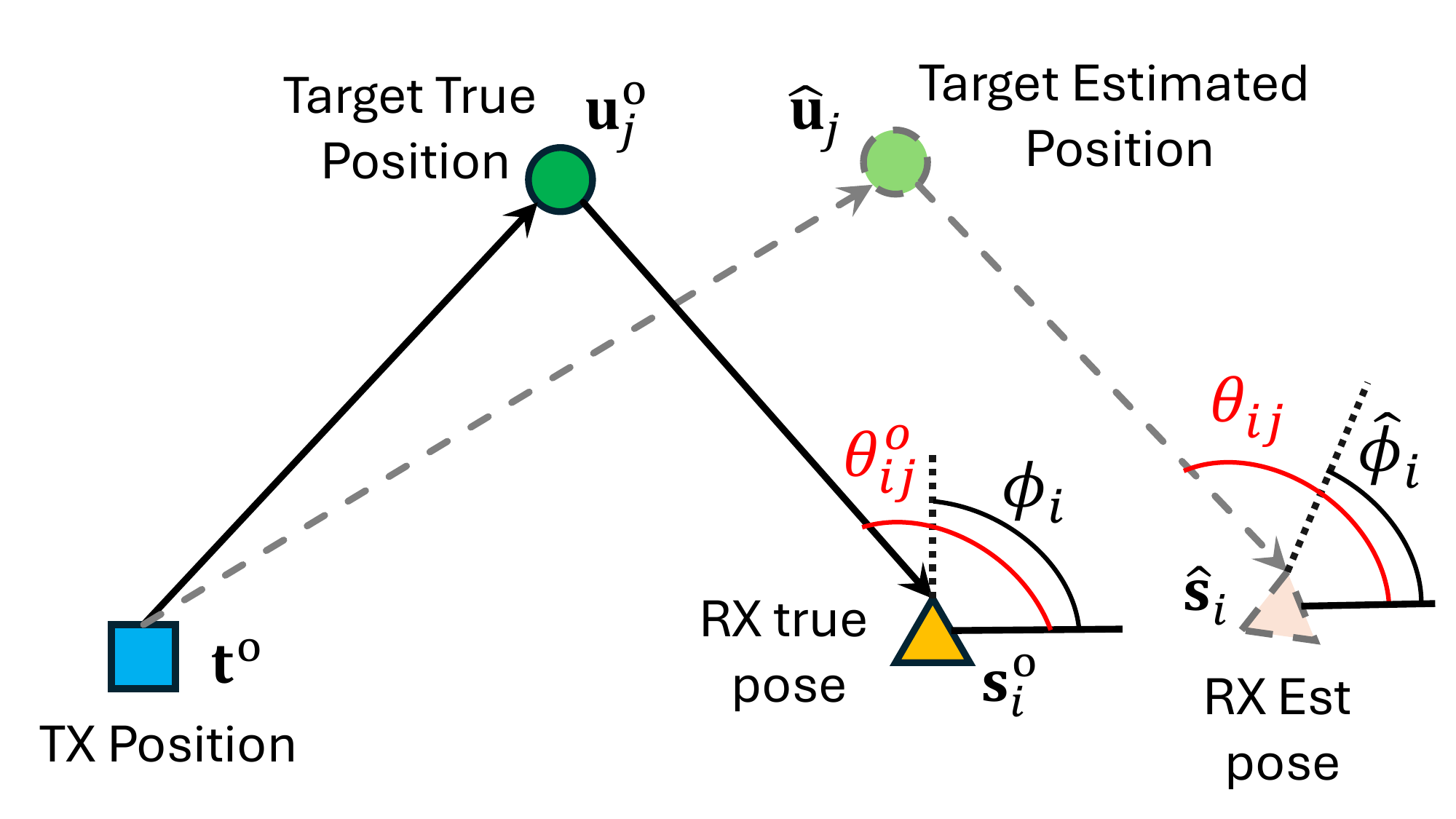}
    \caption{Effect of inaccurate receiver position and orientation on estimated target position, showing ground truth (solid line) and estimate (dashed line).}
    \label{fig:sysmodel}
\end{figure}
Many methods have been proposed to improve the radar pose accuracy for different radar measurement setups. Some works consider the availability of calibrating targets \cite{8827552}, while others rely on cooperation between radar receivers to track a single moving target and improve their own localization \cite{10812006,9139413}. Different methods are used to optimize both the target position and radar pose, including the maximum likelihood (ML) estimator \cite{8827552, 5737800}, which optimizes based on the statistical characteristics of the geometry and measurement model. Other works consider semidefinite programming (SDP) \cite{1599736}, which converts the non-convex optimization problem into a convex one using semidefinite relaxation (SDR). SDP solutions have good performance even under high noise levels, but this comes at the cost of computational complexity.

In this work, we propose a novel approach for joint target localization and receiver pose self-calibration in a multistatic radar system. The proposed approach avoids high complexity by iteratively solving a linear least-squares minimization problem in a closed form. The scheme is designed in a way that allows for a decentralized implementation by exchanging compact local information between receivers.

The main contributions of this work are:
\begin{itemize}
    \item We derive the Cramer-Rao lower bound (CRLB) for a multistatic radar network performing bistatic range and bearing measurements with uncertain receiver poses and known priors.
    \item We formulate an optimization problem to jointly solve for target positions and receiver poses. We then proposed an alternating weighted least-squares algorithm to solve the proposed optimization problem. 
    \item We evaluate the performance of the proposed algorithm through Monte Carlo simulations and demonstrate that it approaches the CRLB under moderate noise levels.
\end{itemize}
The remainder of this paper is organized as follows. Section II introduces the system and measurement models, and formalizes the joint estimation problem. Section III derives the joint CRLB. Section IV presents an alternating estimation algorithm and discusses its implementation details. Section V reports the simulation results. Section VI concludes the paper and outlines directions for future research.

\section{Problem Formulation And Measurement Model}

We consider a multistatic radar in two-dimensional space consisting of a single transmitter at a known location, $M$ bistatic receivers with uncertain positions, and $K$ targets at unknown positions. The goal is to estimate the position of the $K$ targets while simultaneously improving the positioning accuracy of the $M$ receivers. 

Denote the true position of the transmitter by $\mathbf{t}^\circ = [t_x^\circ, t_y^\circ]^\top$. The true position and orientation of the $i^{th}$ receiver are denoted by $\mathbf{s}_i^\circ = [s^\circ_{ix}, s^\circ_{iy}]^\top$ and $\phi^\circ_i$, respectively. The position of the $j^{th}$ target is given by $\mathbf{u}^\circ_j = [u^\circ_{jx},u^\circ_{jy}]^\top$. 

In the considered setup, the receivers' true pose is not known exactly; only an inaccurate estimate obtained through the Global Navigation Satellite System (GNSS) and onboard Inertial Measurement Units (IMU) is available. The inaccurate position and orientation of the $i^{th}$ receiver can be modeled as \cite{9139413}
\begin{equation}
    \mathbf{s}_i = \mathbf{s}_i^\circ + \Delta \mathbf{s}_i, \quad i = 1,2,\dots,M,
\end{equation}
\begin{equation}
    \phi_i = \phi_i^\circ + \Delta \phi_i, \quad  i = 1,2,\dots,M,
\end{equation}
where $\Delta \mathbf{s}_i$ is the position error of the $i^{th}$ receiver, which is modeled as zero-mean Gaussian with a known covariance matrix $\mathbf{Q}_{s}$, and $\Delta \phi_i$ is the orientation error, which follows the Von Mises distribution with a known concentration parameter $\kappa$ corresponding to an angular variance $\sigma^2_{\phi} = 1/\kappa$. For small variance, the Von Mises distribution can be approximated by a zero-mean Gaussian with variance $\sigma^2_{\phi}$. Figure \ref{fig:sysmodel} shows the effect of the inaccuracy in the receiver pose and measurement on the predicted target position.

The receivers are assumed to be synchronized with the transmitter using one of the various methods discussed in the literature \cite{1370671}. Receivers process the received signal that is reflected from the $K$ targets to extract the bistatic range and azimuth angle, the details of which are outside the scope of this paper. The reader can refer to \cite{willis2005bistatic} for further details. The true bistatic range and azimuth angle of the $j^{th}$ target as measured by the $i^{th}$ receiver are given by
\begin{equation}
    r^\circ_{ij} = \|\mathbf{u}^\circ_j - \mathbf{t}^\circ \| + \|\mathbf{u}^\circ_j - \mathbf{s}^\circ_i \|,
\end{equation}
\begin{equation}
    \theta^\circ_{ij} = \operatorname{atan2}(u^\circ_{jy}-s^\circ_{iy},u^\circ_{jx}-s^\circ_{ix}) - \phi^\circ_i,
\end{equation}
where the receiver orientation $\phi^\circ_i$ is subtracted to account for measurements being made in the receiver's local reference frame. The measurements are imperfect and corrupted by noise from various sources. The observed measurements by the receiver in a global frame of reference are thus given by
\begin{equation}
    r_{ij} = r_{ij}^\circ + \Delta r_{ij},
\end{equation}
\begin{equation}
    \theta_{ij} = \theta_{ij}^\circ + \phi_i + \Delta \theta_{ij},
\end{equation}
where $\Delta r_{ij}$ and $\Delta \theta_{ij}$ are additive white Gaussian noise with zero mean and variance, $\sigma_r^2$ and $\sigma_\theta^2$, respectively. 

\section{Cramer-Rao Lower Bound (CRLB)}
In order to evaluate the effect of the receivers' pose error on target localization accuracy and to evaluate the performance of the proposed optimization algorithm, we will derive the CRLB. For ease of presentation, we assume identical position and orientation error variance ($\sigma^2_s$ and $\sigma^2_\phi$) across all receivers and identical bistatic range and bearing measurement variance ($\sigma^2_r$ and $\sigma^2_\theta$) across all the measurement pairs. We will also assume that the measurement and prior noise are independent across receiver and target pairs. The analysis can be easily extended for the case when the variances are different per receiver pose and measurement target-receiver pair.

We begin by defining some vectors that are useful in deriving the CRLB. First, we define the true bistatic range and azimuth angle vectors as 
\begin{equation}
    \mathbf{r}^\circ = [r^{\circ}_{11},\cdots,r^{\circ}_{1K},r^{\circ}_{21},\cdots,r^{\circ}_{MK}]^\top,
\end{equation}
\begin{equation}
    \boldsymbol{\theta}^\circ = [\theta^\circ_{11}, \cdots, \theta^\circ_{1K},\theta^\circ_{21},\cdots,\theta^\circ_{MK}]^\top.
\end{equation}
Hence, the measured range is given by $\mathbf{r} = \mathbf{r}^\circ + \Delta \mathbf{r}$ where $\Delta\mathbf{r}$ is a Gaussian vector with covariance matrix $\mathbf{Q}_r = \mathbf{I}_{MK}\sigma^2_{r}$. Additionally, the measured azimuth angle vector is given by $\boldsymbol{\theta} = \boldsymbol{\theta}^\circ + \boldsymbol{\phi}_i + \Delta \boldsymbol{\theta}$, where $\Delta \boldsymbol{\theta}$ is a zero-mean Gaussian vector with covariance matrix  $\mathbf{Q}_\theta = \mathbf{I}_{MK} \sigma^2_\theta$. Next, we define the receivers' true positions and orientation vectors and the targets' true position vector as
\begin{equation}
    \mathbf{s}^\circ = [\mathbf{s}^{\circ\top}_1,\mathbf{s}^{\circ\top}_2,\dots,\mathbf{s}^{\circ\top}_M]^\top,
\end{equation}
\begin{equation}
    \boldsymbol{\phi}^\circ = [\phi^\circ_1,\phi^\circ_2, \dots, \phi^\circ_M]^\top,
\end{equation}
\begin{equation}
    \mathbf{u}^\circ = [\mathbf{u}^{\circ\top}_1,\mathbf{u}^{\circ\top}_2,\dots,\mathbf{u}^{\circ\top}_K]^\top,
\end{equation}
where we can define the receivers' position and orientation vectors as $\mathbf{s} = \mathbf{s}^\circ + \Delta \mathbf{s}$ and $\boldsymbol{\phi}=\boldsymbol{\phi}^\circ + \Delta \boldsymbol{\phi}$ where $\Delta \mathbf{s}$ and $\Delta \boldsymbol{\phi}$ are zero-mean Gaussian vectors with covariance matrices $\mathbf{Q}_s = \mathbf{I}_{2M}\sigma^2_s$ and $\mathbf{Q}_\phi = \mathbf{I}_M \sigma^2_\phi$, respectively. Finally, we define the stacked measurements and unknown vectors as follows:
\begin{equation}
    \mathbf{m} = [\mathbf{r}^\top, \boldsymbol{\theta}^\top]^\top, 
\end{equation}
\begin{equation}
    \boldsymbol{\alpha}^\circ = [\mathbf{u}^{\circ\top},\boldsymbol{\omega}^{\circ\top}]^\top,
\end{equation}
where $\mathbf{m}$ is a Gaussian vector with covariance matrix $\mathbf{Q}_m = blkdiag(\mathbf{Q}_r,\mathbf{Q}_\theta)$, and $\boldsymbol{\omega}^\circ = [\mathbf{s}^{\circ\top}, \boldsymbol{\phi}^{\circ\top}]^\top$ is the receiver's position and orientation vector, which has a covariance matrix $\mathbf{Q}_\omega = blkdiag(\mathbf{Q}_s, \mathbf{Q}_\phi)$.

We will derive the CRLB for $\boldsymbol{\alpha}^\circ$ under the assumption that the measurement vector $\mathbf{m}$ and receiver position and orientation vector $\boldsymbol{\omega}$ are Gaussian distributed and uncorrelated. The log-likelihood of the probability density function is thus given by,
\begin{equation}
\begin{split}
    \ln f(\mathbf{m},\boldsymbol{\omega};\boldsymbol{\alpha}^\circ)
 = \mathcal{K} & - \frac{1}{2} (\mathbf{m-m}^\circ)^T\mathbf{Q}_m^{-1} (\mathbf{m-m}^\circ) \\ & - \frac{1}{2} (\boldsymbol{\omega}-\boldsymbol{\omega}^\circ)^T\mathbf{Q}_\omega^{-1} (\boldsymbol{\omega}-\boldsymbol{\omega}^\circ),
\end{split}
\end{equation}
where $\mathcal{K} = -1/2ln((2\pi)^{(2MK+3M)}|\mathbf{Q}_m||\mathbf{Q}_\omega|)$ is a constant independent of $\boldsymbol{\alpha}^\circ$. The Fisher Information Matrix (FIM) captures the information content of the measurements and priors about the unknown parameters, and it is given by
\begin{equation}
    \mathrm{FIM}(\alpha^\circ) = E\biggl[ \frac{\partial \ln{f(\mathbf{m},\boldsymbol{\omega} ;\boldsymbol{\alpha^\circ)}}}{\partial \alpha^\circ} \frac{\partial\ln{f(\mathbf{m},\boldsymbol{\omega} ;\boldsymbol{\alpha^\circ)}}}{\partial \alpha^{\circ \top}}\biggr],
\end{equation}
where
\begin{equation}
    \frac{\partial \ln{f(\mathbf{m},\boldsymbol{\omega} ;\boldsymbol{\alpha^\circ)}}}{\partial \alpha^{\circ\top}}  = \biggl[ \frac{\partial \ln{f}}{\partial \mathbf{u}^{\circ\top}},\frac{\partial \ln{f}}{\partial \boldsymbol{\omega}^{\circ\top}} \biggr],
\end{equation}
and the partial derivatives are given by
\begin{equation}
    \frac{\partial \ln{f}}{\partial \mathbf{u}^{\circ\top}} = (\mathbf m - \mathbf m^\circ)^\top \mathbf Q_m^{-1} \frac{\partial \mathbf m}{\mathbf u^{\circ\top}},
\end{equation}
\begin{equation}
    \frac{\partial \ln{f}}{\partial \boldsymbol{\omega}^{\circ\top}} = (\mathbf m - \mathbf m^\circ)^\top \mathbf Q_m^{-1} \frac{\partial \mathbf m}{\boldsymbol{\omega}^{\circ\top}} + (\boldsymbol{\omega} - \boldsymbol{\omega}^\circ)^\top \mathbf Q_\omega^{-1}.
\end{equation}
We define the Jacobian-based information blocks as 
\begin{equation}
    \mathbf{J}_u = \frac{\partial \mathbf{m}}{\partial \mathbf{u}^{\circ\top}} \in \mathbb{R}^{2MK \times 2K}, \quad \mathbf{J}_\omega = \frac{\partial \mathbf{m}}{\partial \boldsymbol{\omega}^{\circ\top}} \in \mathbb{R}^{2MK \times 3M}
\end{equation}
where the Jacobians are derived in Appendix \ref{app:jac}. Then the measurement information blocks are

\begin{equation}
    \begin{matrix}
    \mathbf X= \mathbf J_u^\top \mathbf Q_m^{-1} \mathbf J_u, \\
    \mathbf Y= \mathbf J_u^\top \mathbf Q_m^{-1} \mathbf J_\omega, \\
    \mathbf Z= \mathbf J_\omega^\top \mathbf Q_m^{-1} \mathbf J_\omega.
    \end{matrix}
\end{equation}

Including priors, the full FIM has the following block structure:
\begin{equation}\label{eq:FIM}
\mathrm{FIM}(\boldsymbol \alpha^\circ)=
\begin{bmatrix}
\mathbf X & \mathbf Y\\[4pt]
\mathbf Y^\top & \mathbf Z+ \mathbf Q_\omega^{-1}
\end{bmatrix}.
\end{equation}

Finally, by applying the partitioned matrix inversion formula \cite{petersen2008matrix}, we obtain the CRLB of the target position and node pose, as follows: 
\begin{equation}
    \mathrm{CRLB}(\mathbf u^\circ) = (\mathbf X -\mathbf Y (\mathbf Z+\mathbf Q_\omega^{-1})^{-1}\mathbf Y^\top)^{-1},
\end{equation}
\begin{equation}
    \mathrm{CRLB}(\boldsymbol{\omega}^\circ) = (\mathbf Z+\mathbf Q_\omega^{-1}-\mathbf Y \mathbf X^{-1}\mathbf Y^\top)^{-1}.
\end{equation}

\section{Proposed Algorithm}
In this section, we derive an alternating weighted least squares algorithm that optimizes receiver poses and target positions given the bistatic range and bearing measurements and receiver priors. We first begin by describing the nonlinear minimization problem that jointly optimizes all the unknown parameters. This minimization problem is then broken down into two subproblems, which are solved in an alternating fashion to improve the computational efficiency and allow for decentralized implementation in the future.

\subsection{Joint Optimization}
In order to jointly optimize the target and receiver parameters, we begin by defining the expected measurements based on the current estimate of the unknown parameters as follows, 
\begin{equation}
    \hat{r}_{ij} = \|\hat{\mathbf{u}}_j - \mathbf{t}^\circ\| + \|\hat{\mathbf{u}}_j - \hat{\mathbf{s}}_i \|, \label{rhat_eqn}
\end{equation}
\begin{equation}
    \hat{\theta}_{ij} = \operatorname{atan2}(\hat{u}_{yj}-\hat{s}_{yi}, \hat{u}_{xj}-\hat{s}_{xi}) - \hat{\phi}_i, \label{thetahat_eqn}
\end{equation}
where $\hat{\mathbf{w}}_i = [\hat{\mathbf{s}}_i,\hat{\phi}_i]^\top$ is the  estimated receiver pose, and $\hat{\mathbf{u}}_j$ is the estimated target position. Here, we assume perfect data association throughout, which is achieved by using one of the many methods in the literature \cite{sruti2024non}. 

We define $\Delta \mathbf{m} = (\hat{\mathbf{m}} - \mathbf{m})$ where $\hat{\mathbf{m}} = [\hat{\mathbf{r}}^\top,\hat{\boldsymbol{\theta}}^\top]^\top$ is the expected measurement vector  based on the current estimate of the receiver poses and target positions, and $\mathbf{m} = [\mathbf{r}^\top,\boldsymbol{\theta}^\top]^\top$ is the actual measurement vector. Similarly, we define $\Delta \boldsymbol{\omega} = (\hat{\boldsymbol{\omega}} - \boldsymbol{\omega})$ as the difference between the current receiver pose estimate vector $\hat{\boldsymbol{\omega}} = [\hat{\mathbf{s}}^{\top}, \hat{\boldsymbol{\phi}}^{\top}]^\top$ and the receiver pose estimate prior vector $\boldsymbol{\omega} = [\mathbf{s}^{\top}, \boldsymbol{\phi}^{\top}]^\top$.

Then, the objective function that minimizes the difference between the expected and actual measurements between all receiver and target pairs is defined as,
\begin{equation}
\begin{split}
        \mathbf{V}({\hat{\mathbf{u}},\hat{\boldsymbol{\omega}}})= \frac{1}{2}   \Delta \mathbf{m}^\top \mathbf{W}_m \Delta \mathbf{m} +          \frac{1}{2} \Delta \boldsymbol{\omega}^\top \mathbf{W}_\omega \Delta \boldsymbol{\omega} ,  \label{eq:jointobjective}
\end{split}
\end{equation}
where $\mathbf{W}_m = blkdiag(\mathbf{I}_{MK}\gamma_r, \mathbf{I}_{MK}\gamma_\theta)$ and 
$\mathbf{W}_\omega = blkdiag(\mathbf{I}_{2M}\gamma_s, \mathbf{I}_{M}\gamma_\phi)$ are the weight matrices for the measurements and receiver priors, respectively. The weights $\gamma_r = 1/\sigma_r^2$, $\gamma_\theta = 1/\sigma_\theta^2$, $\gamma_s = 1/\sigma_s^2$, and $\gamma_\phi = 1/\sigma_\phi^2$ are based on the known error variance of the measurements and receiver pose priors.  It can be seen that the objective function is highly non-linear and non-convex due to the non-linearity of the bistatic range and bearing measurements. Additionally, the objective function solves for all unknowns simultaneously, making the optimization inefficient. Therefore, based on the objective function in (\ref{eq:jointobjective}), we will propose an iterative alternating algorithm that reduces computational complexity by breaking down the problem into smaller and easier-to-solve subproblems. 

\subsection{Alternating Least Squares Algorithm}
In order to reduce the computation required, we are going to alternate between updating the receiver poses $\boldsymbol{\omega}$ and target positions $\mathbf u$. This would allow us to turn the optimization into an iterative algorithm, where, in each iteration, we solve two simple optimization problems. Additionally, because we are performing iterative optimization, where we are slowly changing the values of the optimization variables, we can linearize the optimization function and solve it in closed form to further reduce the required computations in each iteration. 

\subsubsection{Subproblem 1 (receiver update)}  

We consider the target positions to be fixed, and update the receiver pose. For each receiver, we create the expected measurements ($\hat r_{ij}$ and $\hat \theta_{ij}$) based on fixed target positions. We then try to minimize the difference between the generated expected measurements and the receiver's actual measurements by optimizing its pose. This is achieved by minimizing the following objective function,
\begin{equation}
\begin{split}
    \mathbf{V}_i(\hat{\boldsymbol{\omega}}_i) =  \frac{1}{2}   \Delta \mathbf{m}_{i}^\top \mathbf{W}_{m,i} \Delta \mathbf{m}_{i}  + 
    \frac{1}{2}\Delta \boldsymbol{\omega}_i^\top \mathbf{W}_{\omega,i} \Delta \boldsymbol{\omega}_i,
\end{split}
\end{equation}
 where $\Delta \mathbf{m}_{i}= \hat{\mathbf{m}_{i}} - \mathbf{m}_{i}$ is a stacked vector of the target observations made by the $i^{th}$ receiver, $\mathbf{W_{m,i}} = \blkdiag(\mathbf{I}_{K_i\times K_i}\gamma_r,\mathbf{I}_{K_i\times K_i}\gamma_\theta) \in \mathbb{R}^{2K_i\times2K_i}$ is the weight matrix of the measurements from the $i^{th}$ receiver, and $\mathbf{W}_{\omega,i} = \blkdiag(\mathbf{I}_2\gamma_s,\gamma_\phi) \in \mathbb{R}^{3\times3}$ is the $i^{th}$ receiver prior weight matrix. 
 
 The objective function can then be linearized by taking the first-order Taylor series expansion around the current estimate of the receiver pose $\hat{\boldsymbol{\omega}}_i$. Let $\delta \boldsymbol{\omega}_i = \hat{\boldsymbol{\omega}}_i - \hat{\boldsymbol{\omega}}_i^{(k)}$ be the update step where $\hat{\boldsymbol{\omega}}_i^{(k)}$ is the current receiver pose estimate at the $k^{th}$ iteration. Then, we can linearize $\hat{\mathbf{m}}_{ij}$ around the current estimate by setting 
\begin{equation}
    \Delta \mathbf{m}_{ij} = \hat{\mathbf{m}}_{ij}^{(k)} - \mathbf{m}_{ij} + \mathbf{J}_{ij} \delta \boldsymbol{\omega}_i = \Delta \mathbf{m}_{ij}^{(k)} + \mathbf{J}_{ij} \delta \boldsymbol{\omega}_i,
\end{equation}
where $\mathbf{J}_i^\omega = \partial \hat{\mathbf{m}}_{ij}/\partial \hat{\boldsymbol{\omega}}_i$ is the Jacobian of the $i^{th}$ receiver measurement vector with respect to the current estimate of the receiver pose provided in Appendix \ref{app:jac}.
Similarly, we define $\Delta \boldsymbol{\omega}_i = \Delta \boldsymbol{\omega}_i^{(k)} + \delta \boldsymbol{\omega}_i$. Then, the linearized objective function is given as
 \begin{equation}
\begin{split}
    \tilde{\mathbf{V}}_i(\delta \boldsymbol{\omega}_i) = & \frac{1}{2} \left(\Delta \mathbf{m}_{i}^{(k)} + \mathbf{J}_{i}^\omega \delta \boldsymbol{\omega}_i\right)^\top \mathbf{W}_{m,i} \left(\Delta \mathbf{m}_{i}^{(k)} + \mathbf{J}_{i}^\omega \delta \boldsymbol{\omega}_i\right) \\
    & + \frac{1}{2}\left(\Delta \boldsymbol{\omega}_i^{(k)} + \delta \boldsymbol{\omega}_i\right)^\top \mathbf{W}_{\omega,i} \left(\Delta \boldsymbol{\omega}_i^{(k)} + \delta \boldsymbol{\omega}_i\right) \label{min_sub1}.
\end{split}
\end{equation}

The objective is now a linear function of the receiver pose estimate update, and the closed-form solution can be obtained by taking the derivative of the objective function in (\ref{min_sub1}) with respect to $\delta \boldsymbol{w}_i$ and setting it to zero. 
\begin{equation}
\begin{split}
    \frac{\partial \tilde{\mathbf{V}}_i}{\partial \delta \boldsymbol{\omega}_i} = \mathbf{J}_{i}^{\omega\top} \mathbf{W}_{m,i} \left(\Delta \mathbf{m}_{i}^{(k)} + \mathbf{J}_{i}^\omega \delta \boldsymbol{\omega}_i\right) +\\ \mathbf{W}_{\omega,i} \left(\Delta \boldsymbol{\omega}_i^{(k)} + \delta \boldsymbol{\omega}_i\right).
\end{split}
\end{equation}

Therefore, the updated receiver pose can be obtained by
\begin{equation}
    \hat{\boldsymbol{\omega}}_i^{(k+1)} =  \hat{\boldsymbol{\omega}}_i^{(k)} - \mathbf{G}_i^{-1} \mathbf{g}_i,
\end{equation}
where 
\begin{equation}
    \mathbf{G}_i = \mathbf{J}_{i}^{\omega\top} \mathbf{W}_{m,i} \mathbf{J}_{i}^\omega + \mathbf{W}_{\omega,i},
\end{equation}
\begin{equation}
    \mathbf{g}_i = \mathbf{J}_{i}^{\omega\top} \mathbf{W}_{m,i} \Delta \mathbf{m}_{i}^{(k)} + \mathbf{W}_{\omega,i} \Delta \boldsymbol{\omega}_i^{(k)}.
\end{equation}

In order to ensure numerical stability of the iterative solution, we will introduce a regularization term $\lambda_i$ that ensures the matrix $\mathbf{G}_i$ is invertible. 
\begin{equation}
    \hat{\boldsymbol{\omega}}_i^{(k+1)} =  \hat{\boldsymbol{\omega}}_i^{(k)} - (\mathbf{G}_i + \mathbf{I}_{3\times3}\lambda_i)^{-1} \mathbf{g}_i. \label{sub1_closed}
\end{equation}

Solving equation (\ref{sub1_closed}) yields the updated pose of the $i^{th}$ receiver in closed form.

\subsubsection{subproblem 2 (target update)}
We consider the receiver poses as fixed, and we optimize each target position. Similar to subproblem 1, this is done by minimizing the following objective function, 
\begin{equation}
\begin{split}
    \mathbf{V}_j(\hat{\mathbf{u}}_j) =  \frac{1}{2}   \Delta \mathbf{m}_{j}^\top \mathbf{W}_{m,j} \Delta \mathbf{m}_{j},
\end{split}
\end{equation}
 where $\Delta \mathbf{m}_j = \hat{\mathbf{m}}_j -\mathbf{m}_j$ is the stacked vector of the difference between the expected and actual measurements of the $j^{th}$ target by all receivers. Here, we are trying to minimize the difference between the expected measurements as a function of the target position and the true measurements. Let $\hat{\mathbf{u}}_j^{(k)}$ be the estimate of the $j^{th}$ target position at the $k^{th}$, and let $\delta \mathbf{u}_j = \hat{\mathbf{u}}_j - \hat{\mathbf{u}}_j^{(k)}$ be the update step. Then we can define the linearized measurement vector around the current estimate as  
\begin{equation}
    \Delta \mathbf{m}_{j} = \hat{\mathbf{m}}_{j}^{(k)} - \mathbf{m}_{j} + \mathbf{J}_{j}^u \delta \mathbf{u}_j = \Delta \mathbf{m}_{j}^{(k)} + \mathbf{J}_{j}^u \delta \mathbf{u}_j,
\end{equation}
where $\mathbf{J}_{j}^u = \partial \hat{\mathbf{m}}_{j}/\partial \hat{\mathbf{u}}_j$ denotes the Jacobian of the $j^{th}$ target measurement vector with respect to its current estimated position as defined in Appendix \ref{app:jac}. The linearized objective function is then given by
\begin{equation}
    \tilde{\mathbf{V}}_j(\delta \mathbf{u}_j) = \frac{1}{2} \left(\Delta \mathbf{m}_{j}^{(k)} + \mathbf{J}_{j}^u \delta \mathbf{u}_j\right)^\top \mathbf{W}_{m,j} \left(\Delta \mathbf{m}_{j}^{(k)} + \mathbf{J}_{j}^u \delta \mathbf{u}_j\right) \label{min_sub2}
\end{equation}
where the objective function is linear, and the closed-form solution can be obtained by taking the derivative of the objective function in (\ref{min_sub2})  with respect to $\delta \mathbf{u}_j$ and setting it equal to zero. 
\begin{equation}
    \frac{\partial \tilde{\mathbf{V}}_j}{\partial \delta \mathbf{u}_j} = \mathbf{J}_{j}^{u \top} \mathbf{W}_{m,j} \left(\Delta \mathbf{m}_{j}^{(k)} + \mathbf{J}_{j}^u \delta \mathbf{u}_j\right).
\end{equation}

Therefore, the updated receiver pose can be obtained by
\begin{equation}
    \hat{\mathbf{u}}_j^{(k+1)} =  \hat{\mathbf{u}}_j^{(k)} - \mathbf{G}_j^{-1} \mathbf{g}_j,
\end{equation}
where 
\begin{equation}
    \mathbf{G}_j = \mathbf{J}_{j}^{u\top} \mathbf{W}_{m,j} \mathbf{J}_{j}^u,
\end{equation}
\begin{equation}
    \mathbf{g}_j = \mathbf{J}_{j}^{u\top} \mathbf{W}_{m,j} \Delta \mathbf{m}_{j}^{(k)}.
\end{equation}

Again, similar to subproblem 1, in order to ensure numerical stability, we will introduce a regularization term $\lambda_j$ which ensures the matrix $\mathbf{G}_j$ is invertible

\begin{equation}
    \hat{\mathbf{u}}_j^{(k+1)} =  \hat{\mathbf{u}}_j^{(k)} - (\mathbf{G}_j + \mathbf{I}_{2\times2}\lambda_j)^{-1} \mathbf{g}_j. \label{sub2_closed}
\end{equation}

The position of the $j^{th}$ target can be updated iteratively by solving the closed-form equation in (\ref{sub2_closed}).

\subsubsection{Joint Alternating Algorithm} The final algorithm is achieved by iterating over subproblems 1 and 2 for each target and receiver until convergence. 

The initial target positions are obtained by solving the ray-ellipse intersection problem \cite{willis2005bistatic} for each receiver-target pair using the known transmitter position $\mathbf{t}^\circ$ and the initial estimate of the receiver pose $\hat{\boldsymbol{\omega}}^{(0)}_i$, and then taking the mean of the position estimates for each target from all receivers. The iterative procedure for jointly optimizing the receiver poses and target positions is presented in Algorithm (\ref{alg:dec-alg}).

Each iteration of the algorithm is $O(MK)$ for the Jacobian computation and $O(M+K)$ for the updates, because the Hessians are $2\times2$ for the target position and $3\times3$ for the receiver pose. Therefore, the complexity is dominated by either Hessian or Jacobian computation, which depends on the number of receivers $M$ and targets $K$. 

Due to linearization and non-convexity of the original objective function, the algorithm will converge to a local minimum that is dependent on the initialization. Thus, it is expected that the algorithm will perform worse for higher prior noise in the receiver poses.

\begin{algorithm}[ht]
\caption{Joint Alternating Algorithm}
\label{alg:dec-alg}
\begin{algorithmic}[1]
    \Require transmitter position $\mathbf{t}^\circ$, receiver's initial estimate $\{\mathbf{s}_i^{(0)},\phi_i^{(0)}\}_{i=1}^M$, measurements $\{r_{ij},\theta_{ij}\}$ $ \forall i \in M$ and $\forall j \in K$, tolerance $\tau_i$ and $\tau_j$, regularization parameters, $\lambda_i$ and $\lambda_j$
    \State $\{\mathbf u^{(0)}_{j}\}_{j=1}^K \gets$ mean target positions based on $\{\mathbf{s}_i^{(0)},\phi_i^{(0)}\}$ and measurements $\{r_{ij},\theta_{ij}\}$.
    \State $k \gets 0$
    \Repeat
    \For{$i \in \{1,\dots,M\}$}
        \State $\hat{\boldsymbol{\omega}}_i^{(k+1)} \gets$ solution of (\ref{sub1_closed})
        \State $\delta \hat{\boldsymbol{\omega}}_i \gets \|\hat{\boldsymbol{\omega}}_i^{(k+1)}-\hat{\boldsymbol{\omega}}_i^{(k)}\|$
    \EndFor
    \For{$j \in \{1,\dots,K\}$}
        \State $\mathbf u_j^{(k+1)} \gets$ solution of (\ref{sub2_closed})
        \State $\delta \hat{\mathbf{u}}_j \gets \|\hat{\mathbf{u}}_j^{(k+1)}-\hat{\mathbf{u}}_j^{(k)}\|$
    \EndFor
    \State $k \gets k+1$
    \Until{$\max_i (\delta \hat{\boldsymbol{\omega}}_i) \leq \tau_i$ and $\max_j (\delta \hat{\mathbf{u}}_j) \le \tau_j$}
    \State \Return $\{\mathbf u^{(k)}_j\}_{j=1}^K$ and $\{\boldsymbol{\omega}^{(k)}_i\}_{i=1}^M$ 
\end{algorithmic}
\end{algorithm}
\section{Simulation Results}
The performance of the proposed algorithm was evaluated using a Monte Carlo (MC) simulation for different geometries. Each experiment was simulated for 1000 trials and the root-mean-square error (RMSE) was computed and compared with the CRLB. Because the CRLB is geometry-dependent, it was computed for each scenario, and then the square root of the mean of the CRLB trace was computed and used for comparison.

The simulation was set up where the transmitter, $M$ receivers, and $K$ targets were uniformly distributed in a $100\times100$ meter area, and the receivers' orientation was uniformly distributed over $[-\pi, \pi]$. We then simulated different parameter values while keeping the remaining parameters constant. The parameters that were considered are node position and orientation variance $\sigma^2_s$ and $\sigma_\phi^2$, bistatic range and bearing measurement variance $\sigma_r^2$ and $\sigma_\theta^2$, and number of receivers $M$ and targets $K$. While performing the MC simulation for one parameter range, the rest of the parameters were fixed at the following values: $M=4$, $K=5$, $\sigma_r = 1 \mathrm m$, $\sigma_s = 5 \mathrm m$, $\sigma_\theta = 1^\circ$, and $\sigma_\phi = 2^\circ$.

\begin{figure}[ht]
    \centering
    \includegraphics[width=1\linewidth]{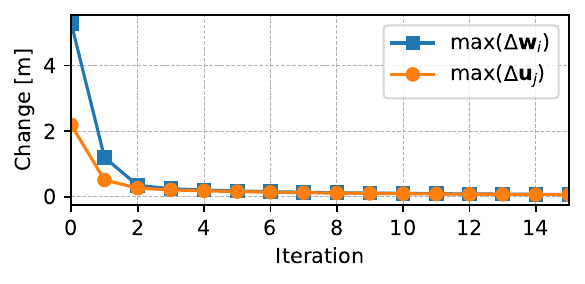}
    \caption{Convergence of the proposed algorithm as shown by change $\boldsymbol{\omega}_i$ and $\mathbf u_j$ per iteration.}
    \label{fig:convergence}
\end{figure}

Figure \ref{fig:convergence} shows the convergence performance of Algorithm \ref{alg:dec-alg} by displaying the convergence criteria, which is the maximum change in $\boldsymbol{\omega}_i$ and $\mathbf u_j$. It can be observed that the algorithm has fast convergence, with the change in position approaching zero after approximately five iterations. 

\begin{figure*}[!t]  
    \centering    \includegraphics[width=0.88\textwidth]{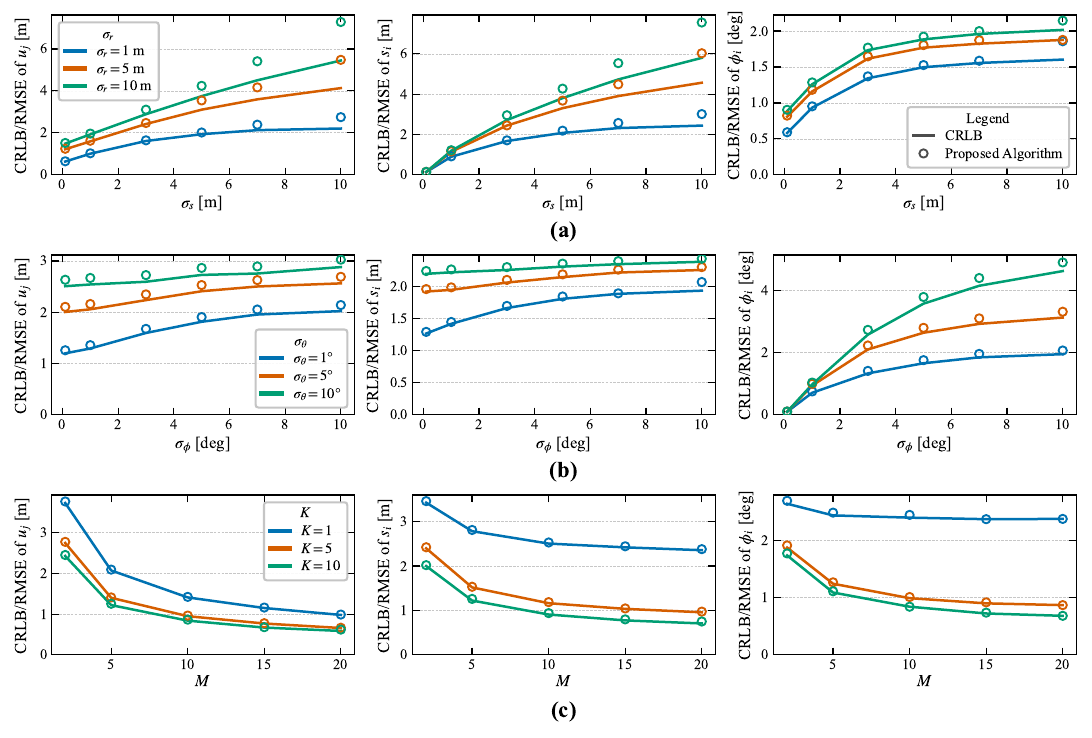} 
    \caption{Parameter sweep results showing CRLB (solid lines) and RMSE of the proposed algorith (open circles) for target position $\mathbf u_j$, receiver position $\mathbf s_i$, and receiver orientation $\phi_i$. In subfigure (a), we see the effect of different values of the receiver position prior noise ($\sigma_s$) and range measurement noise ($\sigma_r$). In (b), we observe the effect of the bearing measurement noise ($\sigma_\theta$) and the receiver prior orientation noise ($\sigma_\phi$). Finally, in (c), we see the effect of the number of receivers ($M$) and the number of targets ($K$).}
    \label{fig:parameter_sweep}
\end{figure*}

Figure \ref{fig:parameter_sweep} (a) shows the RMSE for the target position, node position, and orientation for different values of $\sigma_s$ and  $\sigma_r$. It can be seen that for low to moderate values of both parameters, the performance of the proposed algorithm approaches the CRLB. It is evident that as the noise increases, the performance of the algorithm is degraded because the linearization assumption no longer holds, but it can be observed that even at relatively high noise, the performance is still close to the CRLB. Additionally, the results show that the orientation is less sensitive to changes in $\sigma_r$ and $\sigma_s$.

Figure \ref{fig:parameter_sweep} (b) shows the performance when we varied the noise in the bearing measurement $\sigma_\theta$ and the initial estimate of the receiver orientation $\sigma_\phi$. We see that both the target and receiver position accuracy are not sensitive to changes in $\sigma_\phi$ while the orientation accuracy is very sensitive to changes in both $\sigma_\theta$ and $\sigma_\phi$. Additionally, the results show that the performance of the proposed algorithm approaches the CRLB, and at certain points, it is even below the CRLB, which is due to a slight bias in our estimator caused by the linearization of the highly nonlinear bearing measurement.

Finally, figure \ref{fig:parameter_sweep} (c) shows the performance for different numbers of receivers $M$ and targets $K$. We see that, since we are employing a cooperative algorithm, increasing the number of receivers lowers the error in both target position and receiver pose. However, the improvement diminishes, and after certain values of $M$ and $K$ it becomes ineffective to add more, because it increases the communication overhead and computations required for optimization. 
\section{Conclusions}
This study investigated the joint target localization and receiver self-calibration in multistatic radar systems with uncertain receiver poses. The CRLB analysis revealed that the receiver pose can be updated to be more accurate and to reduce the pose uncertainty below its prior value. A joint optimization non-linear least squares problem was proposed, which was then decoupled into an alternating optimization where we update the receiver poses first and then update the targets' positions. This alternating optimization solves small linearized least squares problems in each iteration, which reduces the computational complexity of the proposed algorithm. Monte Carlo simulations were carried out, which showed that the performance of the proposed algorithm approaches the CRLB for low to moderate noise levels in the measurements and receiver priors. 

\appendices
\section{}\label{app:jac}
Let \(dx=u_x-s_x,\;dy=u_y-s_y,\;d_s^2 = dx^2+dy^2 ,\;d_s=\sqrt{d_s^2},\;d_t=\|u-t\|\). Define
\(\;r_t=(u-t)/d_t,\;r_s=(u-s)/d_s.\;\) Then the row-vector Jacobians are
\[
\begin{aligned}
\mathbf J_{ij}^{r,u} &= (r_t + r_s)^\top, &
\mathbf J_{ij}^{\theta,u} &= \frac{1}{dx^2+dy^2}\big[-dy,\;dx\big],\\
\mathbf J_{ij}^{r,w} &= \big[-r_s^\top,\;0\big], &
\mathbf J_{ij}^{\theta,w} &= \big[-\mathbf J_{ij}^{\theta,u},\;-1\big]=\big[\tfrac{dy}{d_s^2},-\tfrac{dx}{d_s^2},-1\big].
\end{aligned}
\]
Each observed pair \((i,j)\) contributes the 2×(2+3) block,
\(\;J_{ij}=[ \mathbf J_{ij}^{r,u},\mathbf J_{ij}^{r,w};\mathbf J_{ij}^{\theta,u},\mathbf J_{ij}^{\theta,w}]\)
which is placed into \(\mathbf J_u,\mathbf J_\omega\) at the columns for target \(j\) and receiver \(i\).

\bibliographystyle{IEEEtran}
\bibliography{references}
\end{document}